\documentclass[twocolumn]{openjournal}

\usepackage{natbib}
\usepackage{amsmath,amsfonts}
\usepackage{graphicx}
\usepackage[colorlinks,linkcolor=blue,citecolor=blue,urlcolor=blue]{hyperref}

\usepackage[utf8]{inputenc}
\usepackage[T1]{fontenc}
\usepackage{ae,aecompl}

\usepackage{orcidlink}
\usepackage{soul}

\begin{document}


\title{Solar Cycle Irregularity Study}
\shorttitle{Solar Cycle Irregularity Study}

%

 \author{William A. Gardner} 
  \email{bill@cyclicsignals.com}
 \affiliation{Department of Electrical and Computer Engineering, University of California at Davis, \\ 1 Shields Ave, Davis, CA 95616, USA}

 \author{Antonio Napolitano \orcidlink{0000-0003-1457-0349}}
 \email{antonio.napolitano@uniparthenope.it}
 \affiliation{Department of Engineering, University of Napoli ``Parthenope'', \\ Centro Direzionale, Isola C4, 80143 Napoli, Italy \\ \\ }


\label{firstpage}





\begin{abstract}
It has recently been discovered that the time phases (time-varying delays) 
of the irregular periodicities observed in the Sunspot series, consisting of the 
approximate 27-day latitude-averaged plasma rotation and approximate 11-yr magnetic pole reversal, 
share a common pattern or cycle of irregularity. 
Because the method of measurement contains an unknown constant delay of no practical significance, 
the average of the measured cycle of delay is subtracted out, and the resultant cycle proceeds 
from zero delay to maximum delay, back to zero delay, then to maximum advance, and finally back to zero advance/delay. 
The period of this cycle is estimated to be 164 years, which does not allow for even one full repetition of the pattern 
in the 205-yr data record. So, it may or may not be a segment of an ongoing (pseudo) periodicity. 
Nevertheless, the fact that this cycle of irregularity in period is shared by two long-recognized genuine 
irregular periodicities tied to specific activity from two distinct solar phenomena suggests there is a common phenomenon 
affecting these periodicities. 
The physics-based link between the rotation speed of the interior of the Sun and the average over latitude of 
the rotation period of the surface plasma, approximately 27 days on average, suggests the possibility that there may 
be a single primary phenomenon that is responsible for irregularity of both rotation of the interior of the Sun 
and the period, approximately 11 yrs on average, of pole-reversal. 
This is consistent with recently reported (November 2025) results on hypothesized 
planetary forcing of the Solar dynamo. 	
\end{abstract}

\maketitle


\section{Background}
\label{sec:Background}

Paraphrasing from the recent paper 
``On the planetary forcing of the Solar dynamo: Evidence from a Lagrangian framework'' \cite{LeMouel2025}, 
the singular value decomposition of the stationary empirical (time-average) lag covariance matrix is computed jointly 
from two time series, the Sunspot number series and solar Length-of-Day on Earth time series, 
and used to extract their shared pseudo-cyclic components in the form of a mono-modal oscillatory signal 
with evolving amplitude and phase. 
The analysis shows that these two physical phenomena share 12 common components, a trend, and 11 highly correlated pseudo-periods. 
Careful analysis of the phase relationships in the pseudo-cycles associated with the two time series suggests that 
observed coherences are unlikely to be coincidental. 
The 11 pseudo-cycles correspond strictly to the periods found in the sum of the mean 
planetary orbital angular momenta, leading to the hypothesis that these momenta exert a forcing on both 
Sunspot number and Length-of-Day variability, and the conclusion that this is compelling evidence for the existence 
of a unique planetary forcing mechanism acting on the solar dynamo, 
which contributes to the modulation of the solar magnetic cycle. 
Specifically, it is explained that changes in planetary orbital configuration are balanced 
by opposite variations in the Sun's orbital or rotational motion, thereby modulating the Sun's 
rotation rate and, consequently, the solar dynamo. 

The solar dynamo is defined to be the process by which the Sun generates its strong magnetic field through differential 
rotation and convective instability, resulting in the periodic flipping of its magnetic polarity approximately 
every 11 years, known as the solar cycle. 
This process influences the occurrence rate of solar activity phenomena, such as sunspots and coronal mass ejections.

The results in the present article obtained by a different method, which was originally developed for 
analysis of communication signals exhibiting time-varying Doppler, appear to be consistent with the hypothesis 
of a unique forcing mechanism acting on the solar dynamo and might therefore serve as independent corroboration 
of the physical mechanism proposed in \cite{LeMouel2025}.

\newpage

\section{Method}
\label{sec:Method}

This method of modeling and analyzing irregular periodicities in otherwise erratic data is thoroughly 
reviewed in the survey paper \cite{NapolitanoGardner2025} and applied to the 205-year-long record of daily Sunspot series. 
This method begins with the assumption that, were it not for amplitude warping and especially time warping, 
the time series data of interest would exhibit the sum of at least two periodicities corrupted by random fluctuations. 
This method uses knowledge of this data-model structure to estimate the nominal period and the time-warping function 
for each of the two fundamental sine waves found and, when of interest, the amplitude warping function. 
By discovering a time-warping function that is \emph{common to both these pseudo-sine waves} associated with the two 
distinct pseudo periodicities (with pseudo-periods of approximately 27 days and 11 years), 
it is hypothesized that there is a unique cause for the time warping in these two distinct pseudo-periodicities. 
These results provide evidence of a single time-varying delay pattern in the internal rotational behavior 
of the Sun affecting both these periodic phenomena. 

The method calculates the power spectral density of the data and uses a peak-finding search to detect 
two salient pseudo sine waves. The data is then subjected to a bandpass filter centered at each of the peak frequencies, 
and the output of each is angle-demodulated to produce the time-varying delay (or phase) of the pseudo sine wave, 
producing two individual estimates of the time-warping function, and the demodulation frequency used is fine tuned. 
This produces periodicity estimates of approximately 27 days and 11 years. 

If the detected time-warping functions for the two periodicities agree with each other, 
this provides an indication of reliability of the time warp measurement 
(although it is possible in general for different periodicities in data to exhibit different time irregularities, 
depending on the underlying physical phenomenon responsible for the warping). 
This also suggests a new feature characterizing the irregularity of the two well-known periodicities in that data, 
providing evidence that the time warp is a time-varying delay feature reflecting solar physics in contrast to being 
simply an artifact of the data collection, processing, or analysis methods.

\begin{figure}
	\centering
	\includegraphics[width=0.5\textwidth,keepaspectratio]{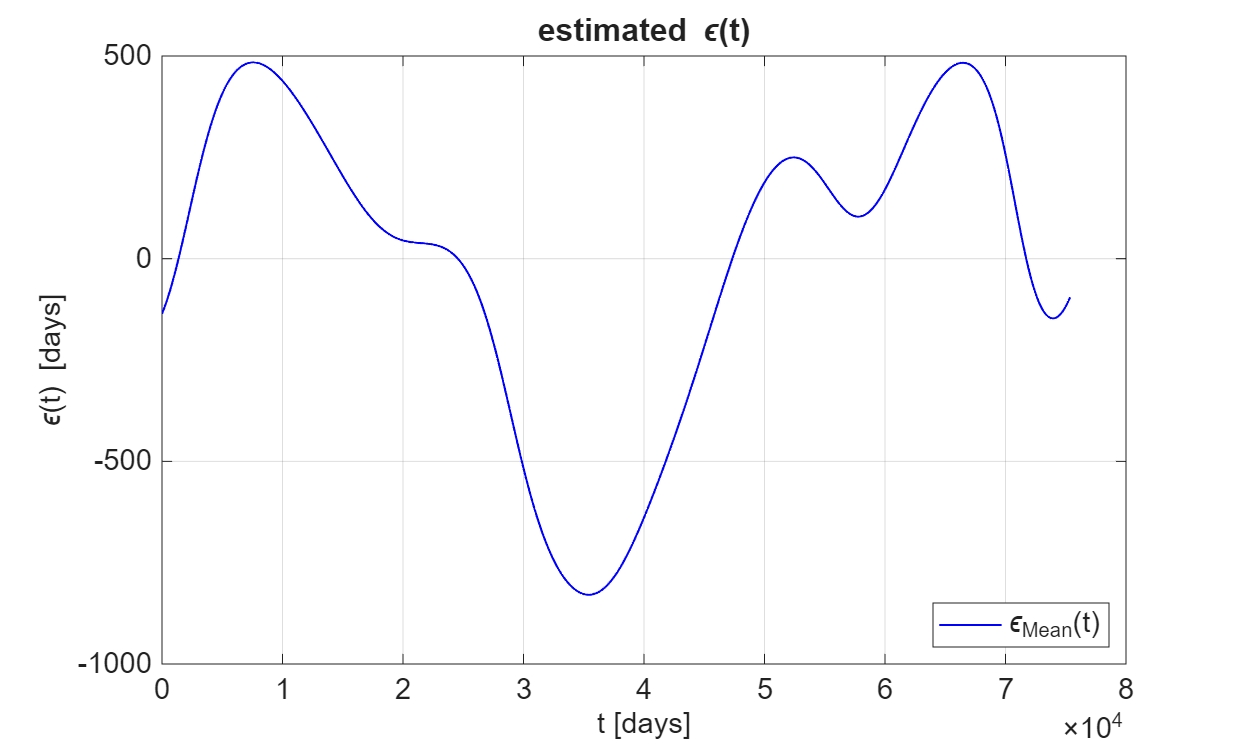}
	\caption{Time delay cycle in approximate 11-year periodicity}
	\label{fig:1}
\end{figure}

\begin{figure}
	\centering
	\includegraphics[width=0.5\textwidth,keepaspectratio]{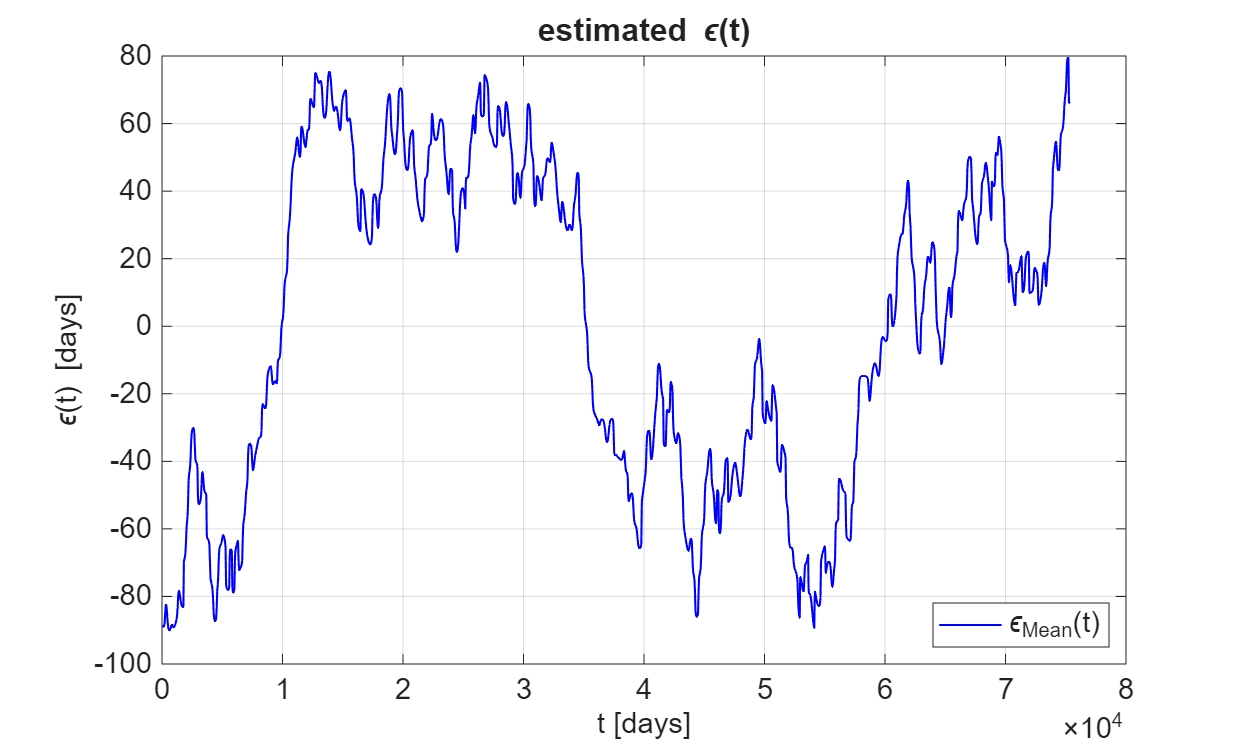}
	\caption{Time delay cycle in approximate 27-day periodicity}
	\label{fig:2}
\end{figure}

This feature roughly approximates a little over one cycle of a sine wave, so estimates for the value of 
the pseudo-period of the feature are obtained by simply making measurements on the graphical display of the data 
shown in the figures. This relatively crude method is deemed acceptable due to the extent of deviation from a true sine wave. 
The feature shown in the curve in Fig.~\ref{fig:1} is a considerably better approximation to a sine wave than 
that shown in Fig.~\ref{fig:2}, which appears to contain substantially more random effects. 

\section{Results}
\label{sec:Results}

As can be seen in Fig.~\ref{fig:1}, the pattern in time delay in the 11-yr periodicity has a dynamic range from maximum 
to minimum of only about 1400 days over about 30,000 days. 
This is equivalent to a change of only about 3.8 yrs in delay over about 82 yrs or 4.6\%. 
For the 27-day periodicity, as can be seen from Fig.~\ref{fig:2}, the dynamic range is about 165 days over about 30,000 days, 
or about 0.55\%. Consequently, although the delay pattern appears similar to a distorted sine wave or a sine wave 
plus low-frequency additive noise for the 11-yr periodicity and, for the 27-day periodicity, it resembles a sine wave 
in quite strong additive broader-band noise (which makes it difficult to say anything about possible distortion), 
there is no way to know from the Sunspot series itself if, 
outside the 205-yr period of observation, the pattern repeats. 
Nevertheless, it is conceivable that the methods and hypotheses in \cite{LeMouel2025} 
could lead to a fit of this feature to the hypothesized modulation of solar orbital or rotational motion.

The approximate period of a single cycle of delay is estimated to be roughly 164 yrs. 
However, this may change over the long run: if this cycle is due to planetary orbital motion, 
as suggested by the work in \cite{LeMouel2025}, then it is probably almost periodic (in the mathematical sense) 
over the long run.
   
Nevertheless, the nature of the difference in dynamic ranges means one periodicity (pole reversal) 
lags behind the other for about 82 yrs and then leads for the next 82 yrs.

\section{Discussion}
\label{sec:Discussion}

One possible explanation for the difference in what is tacitly referred to here as noise level in the measured 
time warp is that the pole reversal phenomenon is less erratic than the plasma surface phenomenon. 
However, there is another possible explanation that attributes the erratic behavior of the measured time delay 
to the measurement method itself. 
Specifically, the demodulation of the phase-modulated 27-day-period sinewave is subject to error when 
the phase variation exceeds 360 deg. This is so even though a phase-unwrapping algorithm is used by the phase demodulator. 
In the case of Fig.~\ref{fig:2}, the large dynamic range of the estimated time-warping function (165 days) 
compared to the period (27 days) means that the sine wave with period 27 days is deeply phase-modulated and, 
consequently, has a large bandwidth. Therefore, the frequency down-conversion and low-pass filtering operations 
\cite[Eq.~(4.2)]{eusipco2024} do not allow one to accurately extract such a single modulated sine wave. 
Specifically, at the output of the low-pass filter, the spectral tails of the modulated 
sine wave of interest are strongly attenuated while spectral tails of other modulated sine waves are present. 
Thus, the angle de-modulation procedure \cite[Eq.~(4.3)]{eusipco2024}
for extracting the time-warping function is not as accurate for the 27-day cycle. 

Future work could use the alternative method based on a projection onto basis functions 
that is presented in \cite{Gardner2018} for estimating the time warp in data that otherwise 
exhibits cyclostationarity with one or multiple periods. 
This method is more computationally intensive than the method used in \cite{NapolitanoGardner2025}  
for the Sunspot Series data, but it is likely to be more accurate, especially if there is information available 
that can inform the choice of basis functions to be used. 
This is the case here because, having shown that the 27-day and 11-year approximate periodicities appear to share 
approximately the same time warp (except what is believed to be noise-like measurement error), 
we can find a good set of basis functions for the time warp shown in Fig.~\ref{fig:1}, 
without using the actual time warp function as a basis function, and then use this basis for 
the method in \cite{Gardner2018} to obtain a better estimate of time warp than that shown in Fig.~\ref{fig:2}.

\section{Conclusion}
\label{sec:Conclusion}

The apparent commonality of the detected cycle of time delay between the 27-day periodicity and the 11-year periodicity, 
except for the dynamic range of the cycle, is in agreement with the proposed single origin of these two periodicities 
presented in \cite{LeMouel2025}, namely planetary forcing of the Solar dynamo \cite{LeMouel2025}. 
Further confidence in the commonality of the time delay cycle may be achievable by obtaining a more accurate estimate 
of the time delay cycle using an alternative method identified in this article, at the expense of higher computational complexity. 
Still further reliability could potentially be obtained using the cyclic covariance 
in addition to the cyclic mean used herein to have two independent estimates of the time delay cycle, 
as done in \cite{NapolitanoGardner2025}.



\bibliographystyle{mnras}
\bibliography{ref} 




\label{lastpage}
\end{document}